\documentstyle[12pt]{article}
\renewcommand{\baselinestretch}{1.0}
\input epsf
\begin{document}

\begin{titlepage}
\hfill TUW--95--24 \\
\hfill gr-qc/9602040 \\

\vfill
\begin{center}

{\LARGE On the Completeness of the Black Hole\\ 
Singularity in 2d Dilaton Theories}\\ 
\vfill
\renewcommand{\baselinestretch}{1}
{\large
{M.O.\ Katanaev\footnote{Permanent address:~Steklov Mathematical Institute, Vavilov 
St., 42, 117966 Moscow, Russia,  \\ katanaev@class.mian.su}}}\\
Erwin Schr\"odinger International Institute \\
for Mathematical Physics\\
Pasteurgasse 6/7, A-1090 Wien\\
Austria\\[0.7cm] 
{\large
{W.\ Kummer\footnote{wkummer@tph.tuwien.ac.at} and H.\ 
{Liebl\footnote{liebl@tph.tuwien.ac.at}}}}\\
Institut f\"ur Theoretische Physik \\
Technische Universit\"at Wien\\
Wiedener Hauptstr. 8-10, A-1040 Wien\\
Austria\\ 
\end{center}
\vfill

\begin{abstract}
The black hole of the widely used ordinary 2d--dilaton model (DBH) 
deviates from the Schwarzschild black hole (SBH) of General 
Relativity in one important feature: Whereas non-null extremals 
or geodesics show 
the expected incompleteness this turns out {\it not to be the case 
for the null extremals}. After a simple analysis 
in Kruskal coordinates 
for singularities with power behavior of this 
-- apparently till now overlooked -- property we discuss the global 
structure of a large family of generalized dilaton theories which does not only
contain the DBH and SBH but also other proposed dilaton 
theories as special cases. For large ranges of the parameters
such theories are found to be free from this defect and exhibit global
SBH behavior.
\end{abstract}
\vfill

Vienna, February 1996 \hfill 
\end{titlepage}
\vfill

\newpage

\section{Introduction}
Dilaton models in 1 + 1 dimensions have been studied extensively in their 
string theory inspired form \cite{wit91}, as well as in generalized versions 
\cite{ban91}. Recently also models with torsion \cite{kat86}
which can be 
motivated as gauge theories for the zweibein have been seen \cite{kat95} to
be locally equivalent to generalized dilaton theories, although the global 
properties differ in a characteristic manner. \newline
The prime motivation for investigating such models and especially 
the ordinary dilaton black hole (DBH), always has been the hope to 
obtain information concerning problems of the 'genuine' 
Schwarzschild black hole (SBH) in d = 4 General Relativity: the 
quantum creation of the SBH and its eventual evanescence because of 
Hawking radiation and the correlated difficulty of information loss 
by the transformation of pure quantum states into mixed ones, black hole 
thermodynamics etc. \cite{alv}. 
On the other hand, essential differences between DBH and SBH  have been
known for a long time. We just quote the Hawking temperature ($T_H$)
and specific heat:
For the DBH $T_H$ only depends on the cosmological constant instead of a 
dependence on the mass parameter as in the SBH.
The specific heat is positive for DBH and negative for the SBH. 

A basis for any application of DBH must be a comparison of its 
singular behavior with the one of the SBH. However, careful 
studies of the singularity structure in such theories seem to be 
scarce. Apart from  \cite{lemos} and our recent work \cite{kat95} we 
are not aware of such a comparison. It always seems to have been 
assumed that the physical features coincide at least qualitatively  
in all respects. During our recent work \cite{kat95} we noted that this is 
not the case: 
{\it For the ordinary dilaton black  hole of
\cite{wit91} 
null extremals are complete at the singularity}. Of course, 
non--null extremals are incomplete, and so at least that property 
holds for the DBH, but, from a physical point of view, it seems a 
strange situation that massive test bodies fall into that 
singularity at a finite proper time whereas it needs an infinite 
value of the affine parameter of the null extremal (describing the 
influx e.g. of massless particles) to arrive. Thus, from the point 
of view of the genuine SBH serious doubts may be raised against any effort 
to extract theoretical insight from the usual DBH. 
\newline
In Section 2 we exhibit the main differences in the presumably most 
direct manner, namely by treating the DBH as well as the SBH as 
special cases of a general power behavior of the metric in Kruskal 
coordinates. We find that the DBH lies at a point {\it just 
outside} the 
end of the interval of the BH-s which are qualitatively equal 
to the SBH in the sense that {\it both} null and non--null 
extremals are incomplete. \newline
In order to pave the way for a more realistic modelling of the SBH 
we then (Section 3) consider a two parameter family of generalized 
dilaton theories which interpolates between the DBH and other models, 
several of whom have been already suggested in the literature 
\cite{lemos,lau,mignemi,fabri}. 
The Eddington--Finkelstein (EF) form of the line element, appearing 
naturally in 2d models when they are expressed as 'Poisson--Sigma 
models' (PSM) \cite{very} is very helpful in this context. 
We indeed find large ranges of parameters for which possibly more
 satisfactory BH models in $d=2$ may be obtained. 

\section{Completeness at a Curvature Singularity}

Consider a metric expressed in Kruskal coordinates {\it u,v}

\begin{equation}
\label{ds}
(ds)^2=2f(uv)dudv \sim 2z^{-a}dudv
\end{equation}
where we assume the metric to be dependent only on the product $uv$ and a simple
leading power behavior of $f(uv)$ near 
\begin{eqnarray}
z=1-uv \to 0. \nonumber
\end{eqnarray}
Without loss of 
generality we consider the space-time where $f>0$ or $uv<1$. 
The case $a=0$ corresponds to
flat Minkowskian space-time and is not considered in the following. This metric
covers many interesting cases. For the DBH we have $f=z^{-1}$ with $a=1$
\cite{wit91}. Neglecting the angular dependence the SBH metric has also this 
form with $a= \frac{1}{2}$.
This can be extracted easily from the formulas on p. 
152 and 153 of ref. \cite{wald}.

From the only nonvanishing components of the affine connection
\begin{equation}
\label{umin}
u^{-1}\Gamma_{11}^1=v^{-1}\Gamma_{00}^0=\frac{d}{d(uv)}\left(\ln f \right)=(\ln f)'
\end{equation}
the curvature scalar becomes at $ z \rightarrow 0$ with (\ref{ds}) 

\begin{equation}
\label{curvdoppel}
R=\frac{2}{f} \left[(\ln f)' + uv(\ln f)'' \right] {
\stackrel{z \to 0}{\longrightarrow} 2az^{a-2}}
\end{equation}

Thus $ a = 2$ (de Sitter behavior) represents  the border between 
singular ($ a <2$) and vanishing ($a >2$) curvature at 
this point. Both SBH and DBH belong to the singular range. With  
(\ref{umin}) the geodesic equations are simply

\begin{equation}
\label{udot}
\ddot{u}+v(\ln f)' \dot{u}^2 =0 ,
\end{equation}
\begin{equation}
\label{udot2}
\ddot{v}+u(\ln f)' \dot{v}^2 =0 ,
\end{equation}
where a dot denotes differentiation with respect to the canonical parameter $\tau.$
The analysis of these equations greatly simplifies by noting that the metric
(\ref{ds}) admits a Killing vector $k^{\mu}, \mu =0,1$, for arbitrary $f$,
\begin{equation}
\label{kill2}
k^{\mu}=\left( \begin{array}{r}
                       -u \\
                        v
\end{array} \right)
\end{equation}
which implies the integral
\begin{equation}
\label{gl7}
\left( \dot{u}v-u\dot{v} \right)f=A_1.
\end{equation}
In the relation between the canonical parameter and the line element
\begin{equation}
\label{a2}
\dot{u} \dot{v} f =A_2
\end{equation}
the constant $A_2>0$, $A_2=0$ and $A_2<0$ describe timelike, null and spacelike
extremals, respectively. The sign of $A_1$ is arbitrary. Expressing $\dot{u}$
and $\dot{v}$ from (\ref{gl7})and (\ref{a2}) in 
$\dot{z}=-u\dot{v} -v\dot{u}$ yields near the singularity
\begin{equation}
\label{zdot}
\dot{z}=\pm \sqrt{A_1^2z^{2a}+4A_2z^a(1-z)}.
\end{equation}
The dependence of $\frac{u}{v}$ on $z$ is determined by
\begin{equation}
\label{ddz}
\frac{d}{dz}\ln \left| \frac{u}{v} \right|=\pm(1-z)^{-1} \left(1+\frac{4A_2}
{A_1^2}(1-z)z^{-a} \right)^{-\frac{1}{2}}
\end{equation}
This follows from (\ref{gl7}) replacing the $\tau$-dependence by a 
$z$-dependence according to (\ref{zdot}). The simple 
analysis of equation (\ref{zdot}) shows that for $0<a<2$ the
curvature singularity can be reached 
at finite affine parameter $\tau$ only by timelike and null extremals,
whereas for $a<0$ this holds for all types of extremals.
The difference between the asymptotic
behavior of null ($A_2=0$) and non-null ($A_2\ne0$) extremals
is especially transparent from (\ref{zdot}).
For null extremals which cross the singularity ($A_1\ne0$)
it takes the form
\begin{eqnarray}                                        \label{ecapag}
  \tau&\sim& z^{(1-a)},~~~~a\ne1,
\\                                                      \label{ecapap}
  \tau&\sim&\ln z,~~~~a=1.
\end{eqnarray}
So null extremals are incomplete (finite value of the canonical
parameter) at the singularity if and only if $a<1$. At the same time
for timelike extremals and $a>0$ equation (\ref{zdot}) yields
\begin{eqnarray} \label{tsim1}
  \tau&\sim& z^{(1-a/2)},~~~~a\ne2,
\\             \label{tsim2}
  \tau&\sim&\ln z,~~~~a=2.
\end{eqnarray}
We see that timelike extremals are incomplete if $0<a<2$. If $a<0$ then
the asymptotic behavior of timelike and spacelike extremals near the
singularity coincides with that of null extremals, (\ref{ecapag}), and
they are always incomplete. Thus we have proved that non-null extremals
are always incomplete at the curvature singularity, $a<2$, $a\ne0$.
Null extremals are incomplete only for $a<1$. So the DBH lies precisely
at the border, $a=1$, where null extremals are still complete at the
singularity. This is a qualitative difference with the SBH at which
all types of extremals are incomplete.

In order to be able to compare the family of dilaton theories 
considered in section 3 below, we also give the transformation of 
(1) into the EF metric
\begin{equation}
\label{ds2}
(ds)^2= d \bar{v}(2d\bar{u}+l(\bar{u})d\bar{v})
\end{equation}

which  explicitly depends on the norm $l = k^\alpha k_\alpha$ of 
the Killing vector $\partial / \partial\bar v$. Introducing   
$F^\prime(y) =f(y)$ the necessary diffeomorphism is ($\sigma = \pm -1$ 
in order to cover the whole original range of $v$)

\begin{eqnarray}
\label{u1}
u=e^{-\bar v}h(\bar u) \\ v=\sigma e^{\bar v}
\end{eqnarray}
with $h(\bar u)$ determined from

\begin{equation}
\label{u2}
\bar{u}=F(\sigma h)
\end{equation}
so that in (\ref{ds2})
\begin{equation}
\label{l1}
l(\bar{u})=-2\sigma h(\bar u) f(\sigma h(\bar u)).
\end{equation}

For the power behavior (\ref{ds}) we obtain
\begin{equation}
\label{l2}
l(\bar{u}) \leadsto |\bar{u}|^{\frac{a}{a-1}}
\end{equation}
to be taken at $\bar u \rightarrow 0_-$ for $a <1$ and at $\bar u \to 
+\infty$ for $a >1$. The DBH case $a = 1$ must be treated 
separately. The result

\begin{equation}
\label{l3}
l_{a=1} \rightarrow e^{\bar u},~~~ \bar{u} \rightarrow \infty,
\end{equation}
as expected, agrees with the familiar DBH behavior \cite{wit91} \cite{alv}.

Eq. (\ref{ds2}) 
is particularly convenient to make contact with the PSM formulation 
which can be obtained for all covariant 2d theories \cite{alt}. They 
may be summarized in a first order action

\begin{equation}
\label{lagr}
L=\int X^+T^- +X^-T^+ +Xd\omega -e^- \wedge e^+V(X)
\end{equation}
In our present case only vanishing torsion

\begin{equation}
\label{tors}
T^{\pm}=(d \pm \omega)  \wedge  e^{\pm}
\end{equation}
as implied by (\ref{lagr}) is expressed in terms of light--cone (LC) 
components for the zweibein one form $e^a$ and for the spin 
connection one form ${\omega^a}_b = {\epsilon^a}_b \omega$ 
where 
 $\epsilon^{ab}=-\epsilon^{ba}$, $\epsilon^{01}=1$ is the
totally antisymmetric tensor. 
The 'potential' V is an arbitrary function of $X$ and determines the dynamics. 
It is simply related to the 
Killing norm $l$ in the EF gauge because (\ref{lagr}) can be solved exactly
for any integrable $V$ 
(cf. the first ref. in \cite{alt}) with the solutions (constant curvature is
excluded)

\begin{eqnarray}
\label{zweib}
e^+=X^+df, \\  \label{zweibb} e^-=\frac{dX}{X^+}+X^-df,
\end{eqnarray}
where $f$ and $X$ are arbitrary functions of the coordinates such that
$df \wedge dX \neq 0$ and $X^+$ is an arbitrary nonzero function.

A similar solution for $\omega$ will not be needed in the following.
The line element immediately yields the EF form (\ref{ds2}) with 
$\bar{u}=\frac{X}{2}$
and $\bar{v}=f$. The Killing norm

\begin{equation}
\label{kill}
l=2\left[ C-w \right] 
\end{equation}
follows from a conservation law  
\begin{equation}
\label{newc}
C=X^+X^- +\int^X V(y)dy=X^+X^- +w
\end{equation}
common to all 2d covariant theories \cite{kat86} \cite{alt} \cite{kat93}
which is related to a 
global nonlinear symmetry \cite{kumwid}. 
The usual dilaton models are produced by the 
introduction of the dilaton field $\phi$ in $X = 2 \exp 
(-2\phi)$, together with a  Weyl 
transformation $e^a=exp(-\phi)\tilde{e}^a$ 
of $d\omega$ in (\ref{lagr})

\begin{equation}
\label{eps}
\epsilon^{\mu \nu} \partial_{\mu} \omega_{\nu} =-\frac{R \sqrt{-g}}{2},
\end{equation}
with the components $\omega_\nu$ expressed by the vanishing of the 
torsion (\ref{tors}) in terms of the zweibein.

The result
\begin{equation}
\label{change}
\sqrt{-g}R=\sqrt{-\tilde{g}}\tilde{R} + 2\partial_{\mu} (\sqrt{-\tilde{g}}
{\tilde g}^{\mu\nu} \partial_{\nu} \phi) 
\end{equation}
will be used in the below and in the next section.

From the preceeding argument there is now a direct relation between 
the singularity in  terms of Kruskal coordinates (\ref{ds}) through 
(\ref{l2}) 
to the singularity in the corresponding action (\ref{lagr}) using 
(\ref{kill})

\begin{equation}
\label{V}
V \rightarrow |X|^{\frac{1}{a-1}}
\end{equation}
with the singularity at $X \to \infty$ for $ 1 < a < 2$ and at $X 
\to 0$ for $a < 1$. The exponential behavior of $V$ for the DBH ($ 
a = 1$) may be read off from (\ref{l3}). 

Let us consider the SBH in a little more detail. The starting point is
the Schwarzschild solution in EF coordinates \cite{wald}

\begin{equation}
\label{ef}
ds^2=2dvdr + \left( 1-\frac{2M}{r} \right)dv^2 -r^2d {\Omega}^2,
\end{equation}
whose $r-v$ part is of the type (\ref{ds2}).
Thus the radial variable
may be identified with $\bar{u}$ in (\ref{ds2}) and with $X$ in (\ref{V}). 
Indeed $a=\frac12$ yields the correct singularity behavior and the corresponding PSM action
(\ref{lagr}) with

\begin{equation}
\label{v2}
V=-\frac{M}{X^2}.
\end{equation}
In the second order formulation (after eliminating the fields $X^{\pm}$, 
and $X$ and using their equations of motion) the action reads
\begin{equation}
\label{r23}
L_{SBH}^{d=2}=\frac{3}{2} \left( \frac{M}{2} \right)^{\frac{1}{3}}
R^{\frac{2}{3}}  \sqrt{-g} .
\end{equation}

Now the metric remains as the only dynamical variable. 
 This action does not look very 
attractive because of the fractional power of curvature,
but in the Euclidean formulation it is positive definite and  
thus may be useful
for quantization. By construction this model beside the flat Minkowskian solution
has the solution of the true SBH.
It should be stressed that the model (\ref{r23}) reproduces the $r,v$-part
of the SBH globally and not only near the singularity.

The above analysis may be extended easily to include also powers of $\log z$ in
(\ref{ds}). For example $f=z^{-1} \ln z$ improves the DBH to make it null
incomplete. In that case $V \to e^{\sqrt{2X}}$ for $X \to \infty$, diverges 
'slightly' less than for the DBH. For exponential behavior in Kruskal coordinates
$f \to e^{-uv}$ for $uv \to \infty$, $R$ diverges exponentially as well. 
In that case both null and non-null extremals are incomplete. Although 
satisfactory in this respect the corresponding PSM potential 
 $V \to \ln (-X) $ at $X \to 0_-$
does not seem particularly useful for applications in BH models.

The 'pure' PSM model  for the SBH with potential (\ref{v2})
is fraught with an important drawback: When 
matter is added the conserved quantity $C$ 
in (\ref{newc}) simply generalizes 
to a similar conserved one with additional matter contributions 
(cf. the second ref. \cite{kumwid}). Thus even before a BH is formed by the 
influx of matter an 'eternal' singularity as given e.g. by (\ref{v2}) for 
the SBH, is present  in which the mass $M$ basically cannot 
be modified by the additional matter. 
A general method to produce  at the same time 
a singularity--free ground 
state with, say, $C = 0$ is provided by a  Weyl 
transformation of the original metric. It simply generalizes what 
is really behind the well-known construction of the DBH theory. 
Consider the transformation 
\begin{equation}
\label{91}
\tilde{g}_{\mu \nu} =\frac{g_{\mu \nu}}{w(X)}
\end{equation}
in (\ref{ds2}) with (\ref{kill}) together with a transformation of $X$

\begin{equation}
\label{92}
\frac{d \tilde{X}}{dX} =w(X(\tilde{X})).
\end{equation}
This reproduces the metric $\tilde{g}_{\mu \nu}$ in EF form
\begin{equation}
\label{93}
(ds)^2=2df \left( d\tilde{X} +\left(\frac{C}{w}-1 \right)df \right)
\end{equation}
with a flat ground--state $C = 0$.  Integrating out
$X^+$ and $ X^-$ 
in (\ref{lagr}), and using the identity  (\ref{change})
with $\phi = \frac{1}{2} \ln w$
one arrives at a generalized dilaton theory

\begin{equation}
\label{95}
\it{L}=\sqrt{-\tilde{g}} \left( \frac{X}{2}R+\frac{Vw}{2}\tilde{g}^{\mu \nu}
\partial_{\mu} \tilde{X} \partial_{\nu} \tilde{X} -Vw \right)
\end{equation}
where $X$ is to be re-expressed by $\tilde X$ through the integral 
of (\ref{92}).  

It should be noted that the (minimal) coupling to matter is 
covariant under this redefinition of fields. Clearly (\ref{95}) is the most 
general action in $d=2$  where the flat ground state
corresponds to $C=0$. $V(X)$ may determine an 
arbitrarily complicated singularity structure. 
The DBH is the special case $V=\lambda^2=const.$ Then $\tilde{X}$ is easily seen
to be proportional to the dilaton field. The SBH results from the choice 
$V=X^{-1/2}.$ Using (\ref{92}) and comparing (\ref{93}) with (\ref{ef})
in that case with the interaction constant in $w$ fixed by 
$w=\frac{\tilde{X}}{2}$,
the conserved quantity $C$ is identified with the mass $M$ of the BH and 
(\ref{95}) turns into the action of spherically reduced 4D general relativity
\cite{lau}. Unfortunately such a theory cannot be solved exactly if coupling
to matter is introduced. Therefore, in the next section a large class of models 
is studied which contain the SBH as a special case.

\section{Dilaton Models Containing Schwarzschild-like Black Holes}

As discussed already in the last section,
any PSM model is locally equivalent to a generalized dilaton model
\cite{kat95}. In the present section we consider all global solutions
of the action

\begin{equation}
\label{ldil}
L=\int d^2x \sqrt{-g}e^{-2\phi}(R+4a(\nabla\phi)^2 +Be^{2(1-a-b)\phi})
\end{equation}
and compare them with the Schwarzschild black hole.
This form of the Lagrangian covers e.g. the CGHS model \cite{wit91}
for $a=1$, $b=0$, spherically 
reduced gravity \cite{lau} $a=\frac{1}{2}$, $b=-\frac{1}{2}$, 
the Jackiw-Teitelboim model \cite{jackiw} $a=0$, $b=1$.
Lemos and Sa \cite{lemos} give the global 
solutions for $b=1-a$ and all values of $a$,  
Mignemi \cite{mignemi} considers
$a=1$ and all values of $b$. The models of \cite{fabri} correspond to $b = 
0, a \leq 1$. The different models can be arranged in an $a$ vs. $b$ 
diagram (Fig.1).

The Lagrangian (\ref{ldil}) is obtained from (\ref{lagr}) by the "generalized 
dilatonization",  as explained in the last section with $\phi$ replaced
by $a\phi$ in (\ref{change})

\begin{equation}
\label{trans}
e^d_\mu =e^{-a\phi} \tilde{e}^d_\mu ~~~~\Leftrightarrow ~~~~
g_{\mu\nu}= e^{-2a\phi} \tilde{g}_{\mu\nu}, 
\end{equation}
where $a$ is an arbitrary constant.
We also replace $X$ 
by the usual dilaton field

\begin{equation}
\label{x}
X=2e^{-2\phi} 
\end{equation} 
and restrict ourselves to a power behavior in $V$
\begin{equation}
\label{v}
V(X)=B \left( \frac{X}{2} \right)^b, 
\end{equation}
depending on the parameters $b$ and $B$ \cite{mignemi2}.
Using the general solution (\ref{zweib}),(\ref{zweibb}) 
and defining coordinates $v=-4f$,
$u=\phi$ immediately yields the line element corresponding to (\ref{ldil})

\begin{equation}
\label{metric}
(d \tilde{s})^2= g(\phi) \left( 2dvd\phi+ \tilde{l}(\phi) dv^2 \right),
\end{equation}
with
\begin{eqnarray}
\label{Gl:34}
b \neq -1: & \tilde{l}(\phi)=\frac{e^{2\phi}}{8} \left(C- \frac{2B}{b+1} e^{-2(b+1)\phi} \right), \\
\label{Gl:34b}
b =-1: & \tilde{l}(\phi)=\frac{e^{2\phi}}{8} \left( \tilde{C}+4B\phi \right),~~~\tilde{C}=C-2B \ln 2 
\end{eqnarray}

\begin{equation}
g(\phi)= e^{-2(1-a)\phi}
\end{equation}
where $C$ is the arbitrary constant defined in (\ref{newc}).
Calculations can be performed in this form or by changing to a new variable

\begin{eqnarray}
\label{Gl:34c}
a\neq1: &u=\frac{e^{-2(1-a)\phi}}{2(a-1)}, \\ a=1: &u=\phi,
\end{eqnarray}
to obtain the EF form (\ref{ds2}) of the metric with

\begin{eqnarray}
\label{newl}
b\neq -1: & a\neq 1: & l(u)=B_1|u|^{\frac{a}{a-1}} -B_2|u|^{1-\frac{b}{a-1}}, \\
 & a=1: & l(u)=\frac18 e^{2u}\left(C-\frac{2B}{b+1}e^{-2(b+1)u} \right), \\
b=-1: & a \neq 1: & l(u)=\frac18|2(a-1)u|^\frac a{a-1}
  \left(C+\frac{2B}{a-1}\ln|2(a-1)u|\right), \\
 & a=1: & l(u)=\frac{e^{2u}}{8} \left( \tilde{C} +4Bu \right),
\end{eqnarray}
where the constants are given by
\begin{equation}
\label{a1}
B_1=\frac{C}{8} (2|a-1|)^{\frac{a}{a-1}}~~~~B_2 = \frac{B}{4(b+1)}(2|a-1|)^
{1-\frac{b}{a-1}}.
\end{equation}
The range of $u$ for $a=1$ is $[-\infty, +\infty]$
whereas from (\ref{Gl:34c}) one has
$$
  2(a-1)u>0,
$$
and for $a>1$ the range is reduced to $[0,+\infty]$ and for $a<1$ it
is $[-\infty,0]$.
Now the scalar curvature in the EF coordinates
$$
  R=l,_{uu}
$$
can be easily calculated
\begin{eqnarray}                                        \label{curv}
b\ne-1, & a\ne1: & R=B'_1u^{\frac{2-a}{a-1}}+B'_2u^{\frac{a+b-1}{1-a}}
\\                                                      \label{ecnboa}
b\ne-1, & a=1: & R=\frac C2e^{2u}-\frac{Bb^2}{b+1}e^{-2bu}
\\                                                      \label{ecobna}
b=-1, & a\ne1: & R=\frac a2|2(a-1)u|^{\frac{2-a}{a-1}}\left(C+\frac{2B}a+2B
  +\frac{2B}{a-1}\ln|2(a-1)u|\right)
\\                                                      \label{ecoboa}
b=-1, & a=1: & R=\frac12e^{2u}(\tilde{C}+4B+4Bu).
\end{eqnarray}
with $B'_1$ and $B'_2$ to be determined easily from (\ref{a1}).
In the following we will continue our analysis in terms of $\phi$
which can always be transformed to $u$ by means of (\ref{Gl:34c}). 
We see that depending on the values of $a$ and $b$ the scalar curvature
may be singular in the limits $u\rightarrow0,\pm\infty$. In Fig.2
the dashed regions show singularities of the scalar curvature.
In the limit $\phi\rightarrow\infty$ the singular regions are different
for zero and nonzero value of $C$,
\begin{eqnarray}                                        \label{ecuscn}
  \phi &\rightarrow& +\infty,~~C\ne0,~~|R|\rightarrow\infty,~~
  (a<2)\cup(a+b<1)\cup(a=2,b=-1),
\\                                                      \label{ecuscz}
  \phi &\rightarrow& +\infty,~~C=0,~~|R|\rightarrow\infty,~~
  (a<1)\cup(a+b<1)\cup(a=2,b=-1).
\end{eqnarray}
The singularity of $R$ may be positive or negative depending on the
values of the constants $a$, $b$, $B$, and $C$ as can be easily seen
from (\ref{curv})--(\ref{ecoboa}). In the limit $\phi\rightarrow-\infty$
the scalar curvature is singular when
\begin{equation}                                        \label{ecuspn}
  \phi \rightarrow -\infty,~~\forall C,~~|R|\rightarrow\infty,~~
  (a+b>1)\cup(a>2)\cup(a=2,b=-1).
\end{equation}

Now one has to analyse the behavior and completeness of extremals
corresponding to the line element (\ref{metric}).
For the first term in (\ref{newl}) the behavior at the singularity
for $|u|\to0$, $a<1$ coincides with (\ref{l2}) if $u=\bar{u}$ and
if $a$ denotes the same parameter as in section 2.
However for a general discussion of (\ref{newl})
the behavior of both terms and the zeros of $l$ (horizons) are better
discussed in terms of the general procedure as outlined in \cite{alt}
of which a summary also was given in \cite{kat95}.
The equations for extremals for the EF metric read $(l'=dl/du)$
\begin{eqnarray}                                        \label{37}
  \ddot u+l'\dot u\dot v+\frac12ll'\dot v^2&=&0,
\\                                                      \label{38}
  \ddot v-\frac12l'\dot v^2&=&0.
\end{eqnarray}
Thus the null extremals $v_1=const$ are always complete 
because $\ddot u =0$ follows from (\ref{37}) and (\ref{38}).
The second
null direction
\begin{equation}
\label{second}
\frac{dv_2}{du}=-\frac{2}{l}
\end{equation}
inserted into (\ref{37}) also yields $\ddot u=0$ or equivalently
$$
  d\tau \sim e^{2(a-1)\phi}d\phi.
$$
Thus the completeness of null extremals depends only on $a$:
\begin{eqnarray}                                        \label{econep}
  \phi &\rightarrow& +\infty,~~a\ge1,~~{\rm complete},
\\                                                      \label{econem}
  \phi &\rightarrow& -\infty,~~a\le1,~~{\rm complete},
\end{eqnarray}
as shown in Fig.3. The physically 
unreasonable behavior of null extremals in this case, encountered already
in section 2, thus has been reconfirmed also here.

The affine  parameter of non-null extremals is obtained 
from an integral similar to (\ref{gl7}) 

\begin{equation}
\label{Gl:53}
\frac{du}{d\tau} + l\frac{dv}{d\tau} = \sqrt{A} = const.
\end{equation}
Identifying the affine parameter $d\tau$ with the $ds$ in 
(\ref{ds2}), these extremals are found to obey 

\begin{equation}
\label{Gl:54}
\frac{dv}{du} = -\frac1l \left[ 1 \mp ( 1 - l /A)^{-1/2}\right] 
\end{equation}
by simply solving a quadratic equation. In addition,  from 
(\ref{Gl:53}) and (\ref{Gl:54}) the affine parameter is determined by
\begin{equation}
\label{Gl:55}
\tau(u)= \int^{u} \frac{dy}{\sqrt{A-l(u)}} ~~~ .
\end{equation}

For $A > 0$, resp. $ A < 0$ the parameter $s$ is a timelike, resp. 
spacelike quantity. 
As for the null extremals our results are given in terms of $\phi$. Non null
extremals turn out to be complete for
\begin{eqnarray}                                        \label{econnm}
\phi \rightarrow-\infty,&~~(a\le1) \cap (a+b\le1), \\          \label{ecuscnn}
\phi \rightarrow+\infty,&~~C\ne0, (a \geq2 )\cap(a+b\geq1),
\\                           \label{ecusczz}
\phi \rightarrow+\infty,&~~C=0, (a\geq1)\cap(a+b\geq1).
\end{eqnarray}
as depicted in Fig.3b,c. 
One observes that only for $\phi \to +\infty$
the region of incompleteness coincides with the one for $|R| \to \infty$
except for the point $a=2,b=-1$ where non null extremals are complete at
the curvature singularity.
As depicted in Fig.2 we find for some distinct values of $a$ and $b$ de
Sitter space or flat space-time solutions:
\begin{eqnarray}
\label{desit}
de Sitter: & a=2, b=0,1~~~a=0,b=1 \\
flat: & a=0, b=0
\end{eqnarray}
Furthermore we get the special regions:
\begin{eqnarray}
\label{newregion}
R=C:&a=2 ~~~\cap ~~~B=0 \\
R=0:&B=0 ~~~\cap ~~~C=0 \\
R=-2\frac{Bb^2}{b+1}:&(C=0)\cap (a+b=1) \cap (B\neq0) \cap (b\neq 0)
\end{eqnarray}

Now elegant methods \cite{alt} exist to find the global structure of 2D models.
Equation (\ref{second}) determines the shape for a certain patch, e.g. the 
typical one for a black hole drawn in Fig.4a. 
The fully extended global solution
across the dotted lines is obtained by gluing together patches of this type
identifying lines with $u=const.$ and exchanging the two null directions
\cite{alt} by an (always existing) diffeomorphism in the overlapping square
or triangle. This simply amounts to using reflected or rotated  building blocks
of the same structure, so as to arrive e.g. for the BH at its characteristic
Penrose diagram Fig.4b.
For definiteness
we consider the case $B\ge0$. The case $B<0$ is obtained by rotating the
diagrams for $B>0$ by 90 degrees, because formally both cases
are related to each other by exchanging the coordinates of
space and time.
Up to a rotation there are six types of Penrose diagrams D1,\dots,D6,
as shown in Fig.5. The boundaries of the diagrams correspond to infinite
values of the dilaton field $\phi=\pm\infty$ as indicated. There the
scalar curvature may be singular
or nonsingular. For the moment we concentrate on the shape of the
diagrams which is defined entirely by the function $\tilde l$ (\ref{Gl:34}) or
(\ref{Gl:34b}). Therefore it does not depend on the value of $a$.
We summarize the types of Penrose diagrams corresponding to
different values of the constants in the following table:

\underline{$B>0$}
\begin{equation} \label{tablbg}
\begin{array}{l}
D1:~~~C=0,~b>0;   \\
D2:~~~C<0,~-1<b\le0;~~~C=0,~(b<-1)\cup(-1<b<0);~~~C>0,~b<-1;  \\
D3:~~~C<0,~b>0;  \\
D4:~~~C>0,~b>0;  \\
D5:~~~C<0,~b\le-1;~~~C=0,~b=-1;~~~C>0,~-1\le b\le0;  \\
D6:~~~C=0,~b=0.
\end{array}
\end{equation}
For $B=0$, $C\ne0$ a global solution is of the type D2,
whereas for $B=C=0$ one has a flat solution.
Using (\ref{curv})--(\ref{ecuspn}), or equivalently Fig.2
it is straightforward to verify
where boundaries are singular, asymptotically flat or correspond to
constant curvature. Completeness at the boundaries follows immediately 
from (\ref{econep})--(\ref{econnm}).

In order to admit a Schwarzschild
like global solution, the Penrose diagram must be of the type D5 with
incomplete spacelike singular boundary $\phi\rightarrow+\infty$ and
complete asymptotically flat boundary $\phi\rightarrow-\infty$.
The above analysis proves that this happens if and only if
the parameters satisfy the following conditions
\begin{equation} \label{tabscs}
\begin{array}{ll}
  B>0,~~a<1, & C<0,~~~b\le-1,   \\
             & C=0,~~~b=-1,     \\
             & C>0,~~~-1\le b\le0.
\end{array}
\end{equation}
In Fig.6 we have tried to summarize all possible cases with one horizon.
The range of (\ref{tabscs}) corresponds to the lower left corner $a<1,~b\leq 0.$
 
Known soluble models with couplings to scalar matter 
\cite{wit91,ban91,lemos,fabri}    
always 
start from an 'undilatonized' (PSM--type) action with $V = 
\lambda^2$, i.e. from a flat theory in which scalar fields are 
introduced. The subsequent dilatonization moves the singularity to 
the factor of $C$, as  explained in the previous section 
and $B_1 = C = 0$ should be a Minkowskian space-time. 

We want to stress again that in the presence of matter it is this factor
(the 'mass' parameter) which is changed by the influx of matter, starting
say with a ground state $C=0$.
The only possibility to start with a Minkowskian space-time before matter flows
in $(C=0 \to C\neq 0)$ is
the case when the $B_2$-term of (\ref{newl}) looses its $u$
dependence completely, which implies that $b=a-1$. 
From Fig.6 we see immediately that 
the CGHS model as well as spherically reduced 4D gravity lie on that line.
If one is interested to reproduce a SBH like model this means that one has
to start out with an action of the restricted form
\begin{equation}
\label{future}
L=\int d^2x \sqrt{-g}e^{-2\phi}(R+4a(\nabla\phi)^2 +Be^{4(1-a)\phi}),~~~a<1
\end{equation}
Similarly we see that for $b=0$ one inevitably obtains Rindler space-time 
\cite{fabri} except
for the CGHS model $(a=1)$.   

The soluble models with matter couplings 
(corresponding to $b = 0$ \cite{wit91,fabri}) 
are compatible with $b=a-1$ only at $a=1$, i.e. for the DBH. 
Although the curvature vanishes for the asymptotic Rindler like space-times
in these models
this could be interpreted as 
observing a black hole formation and its Hawking radiation from an 
accelerating frame --- which does not seem very satisfactory.



It may be added that also the problem of a mass independent Hawking temperature
$\kappa$ appears for $b=0$ models. We take its geometric definition
\cite{wald} from the norm of the Killing vector $k^2=l$ at the horizon,
determined by $l(u_h)=0$,
\begin{equation}
\label{hawktemp1}
\partial_{\mu} k^2 \arrowvert_{u_h} = 2 \kappa k_{\mu} \arrowvert_{u_h}
\end{equation}
i.e.
\begin{equation}
\label{hawktemp2}
2 \kappa =l'(u_h)
\end{equation}

Using (\ref{newl}) one easily finds the Hawking temperature
\begin{eqnarray}
\label{hawktemp3}
  b\ne-1:~~~~
  2\kappa&=&\mp\frac{b+1}{1-a}|B_1|^{\frac b{1+b}}|B_2|^{\frac1{1+b}},
\\
\label{hawktemp4}
  b=-1:~~~~
  2\kappa&=&\frac B2e^{-\frac C{2B}},
\end{eqnarray}
where minus and plus signs in eq.(\ref{hawktemp3}) correspond to the
cases $B_1>0$, $B_2>0$ and $B_1<0$, $B_2<0$ respectively. In the case
$B_1B_2<0$ the black hole solution is absent as follows from (\ref{tabscs}).
We see that the models with $b=0$ share the unpleasant feature with
the $DBH$ that $\kappa$ is then independent of the conserved quantity $C$
interpreted as the mass since $B_1$ disappears.
Another interesting consequence of eqs.(\ref{hawktemp3}), (\ref{hawktemp4})
is the restriction on the coupling constants following
from the positiveness of the temperature: only the first two cases in
(\ref{tabscs}) survive.

\section{Summary and Outlook}

Because systematic  investigations of the global properties of BH 
models are still relatively rare, an important defect of the 
ordinary dilaton black hole relative to 
the genuine Schwarzschild one seems to have been overlooked until now: 
Its singularity is only incomplete with respect to {\it non--null} 
extremals. The special role of the DBH has been put into perspective 
by first embedding it into a family of singularities, to be analyzed 
according to a power behavior in Kruskal coordinates near the singularity. 
We also found a large class of models which comprises well known theories
and which possesses BHs for a certain range of the parameters. Demanding
furthermore that for $C=0$ (no matter)  we have Minkowski space reduces
the allowed region to a straigth line $b=a-1$ (cf.  Fig.6). Not surprisingly
spherically reduced 4D gravity corresponds to one point on that line 
$(a=\frac{1}{2}),$ as well as the DBH $(a=1)$. For exactly solvable models
with matter, withtin this class of theories the asymptotic 
background inevitably is of Rindler type and thus may 
lead to problems of interpretation. In our opinion
a soluble model (after interaction
with scalar matter is added) with qualitative features coinciding completely
with the SBH still waits to be discovered.

\section*{Acknowledgement}

We are grateful for discussions with 
H. Balasin, T. Kl\"osch, S. Lau and M. Nikbakht.
This work has been supported by Fonds zur F\"orderung der
wissenschaftlichen For\-schung (FWF) Project No.\ P 10221--PHY.  One
of the authors (M.K.) thanks The International Science Foundation,
Grant NFR000, and the Russian Fund of Fundamental Investigations,
Grant RFFI--93--011--140, for financial support.

\newpage

\begin{center}
\leavevmode
\epsfxsize=10cm
\epsfbox{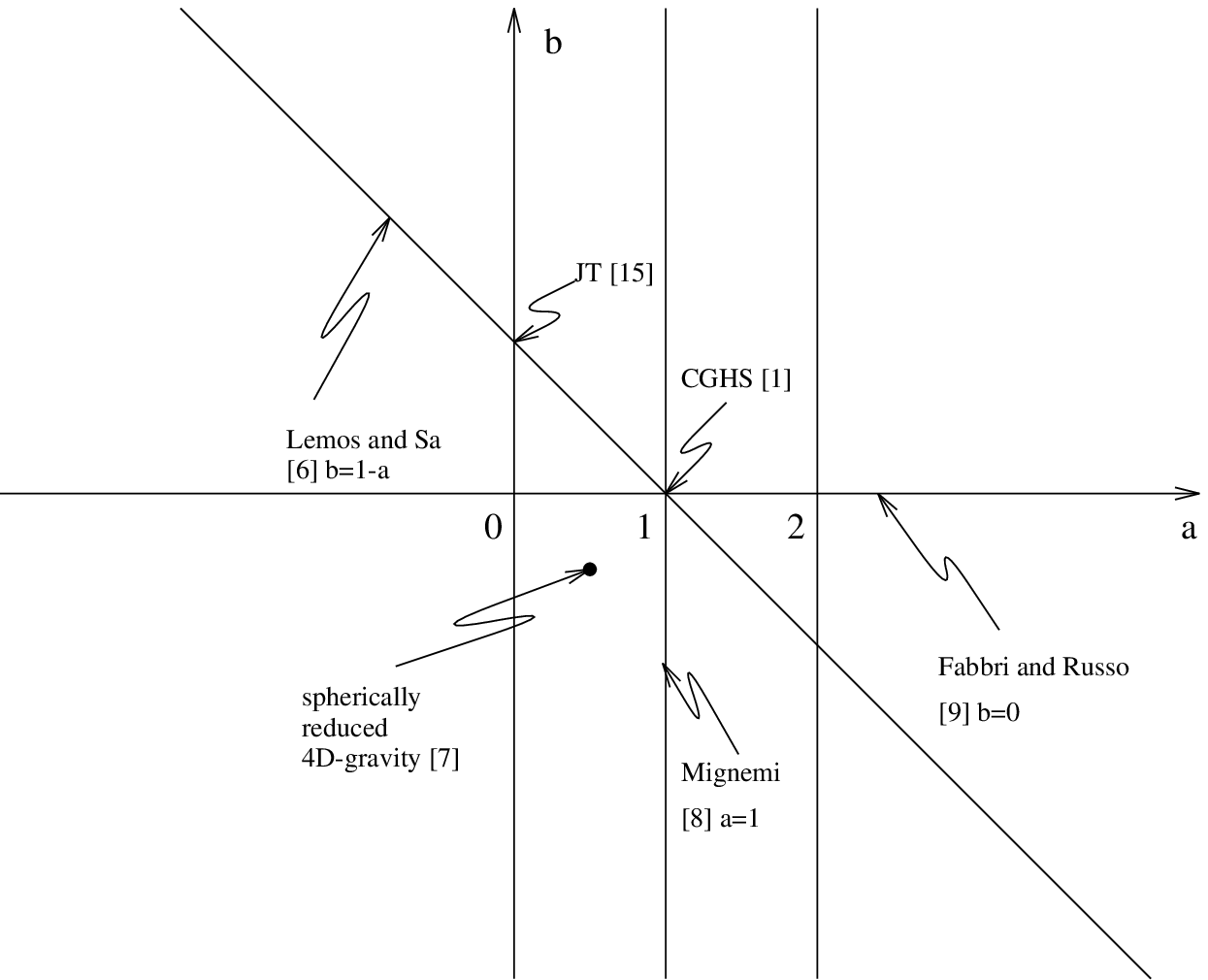}
\\
\centerline{Fig.\refstepcounter{figure}$\;$ \thefigure
\smallskip {}~~Different models in the a-b parameter plane}
\end{center}

\begin{center}
\leavevmode
\epsfxsize=15cm
\epsfbox{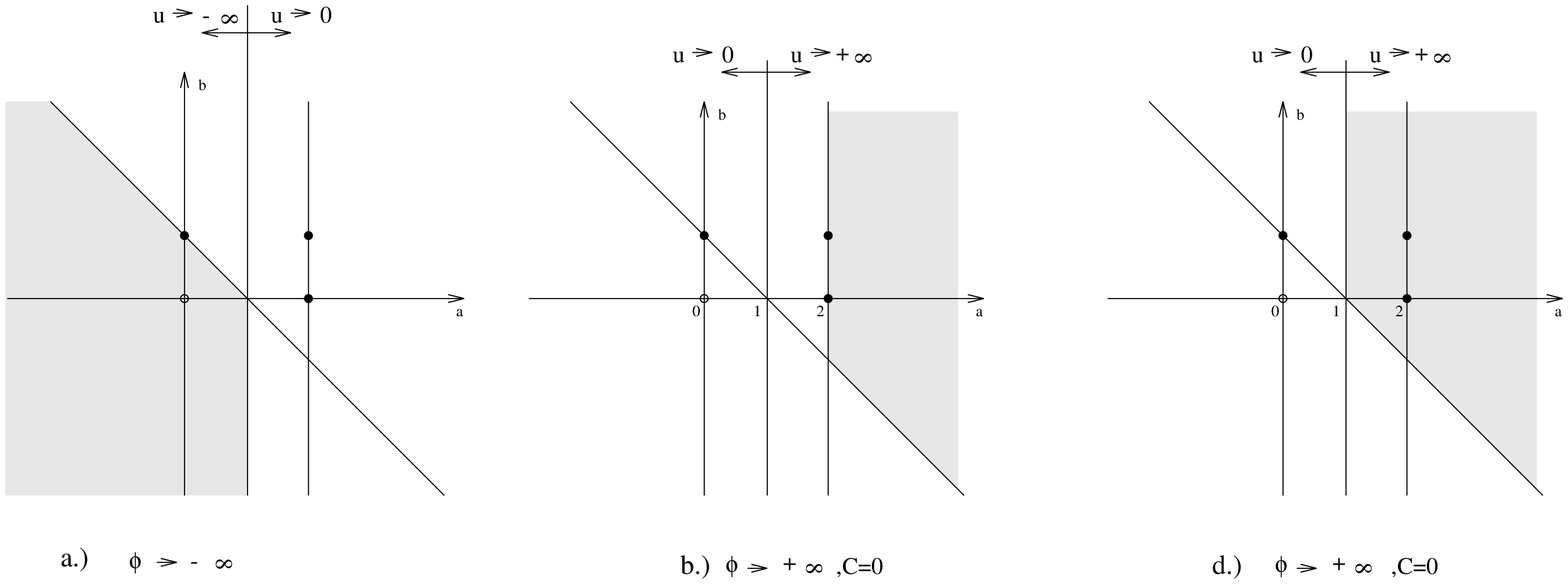}
\\
{Fig.\refstepcounter{figure}$\;$ \thefigure
\smallskip {}~~~Unshaded regions correspond to curvature singularities
as $|\phi | \to \infty$. The boundaries belong to the shaded region. 
Full dots show de Sitter space-times whereas empty dots indicate flat 
space-times.}
\end{center}

\begin{center}
\leavevmode
\epsfxsize=14cm
\epsfbox{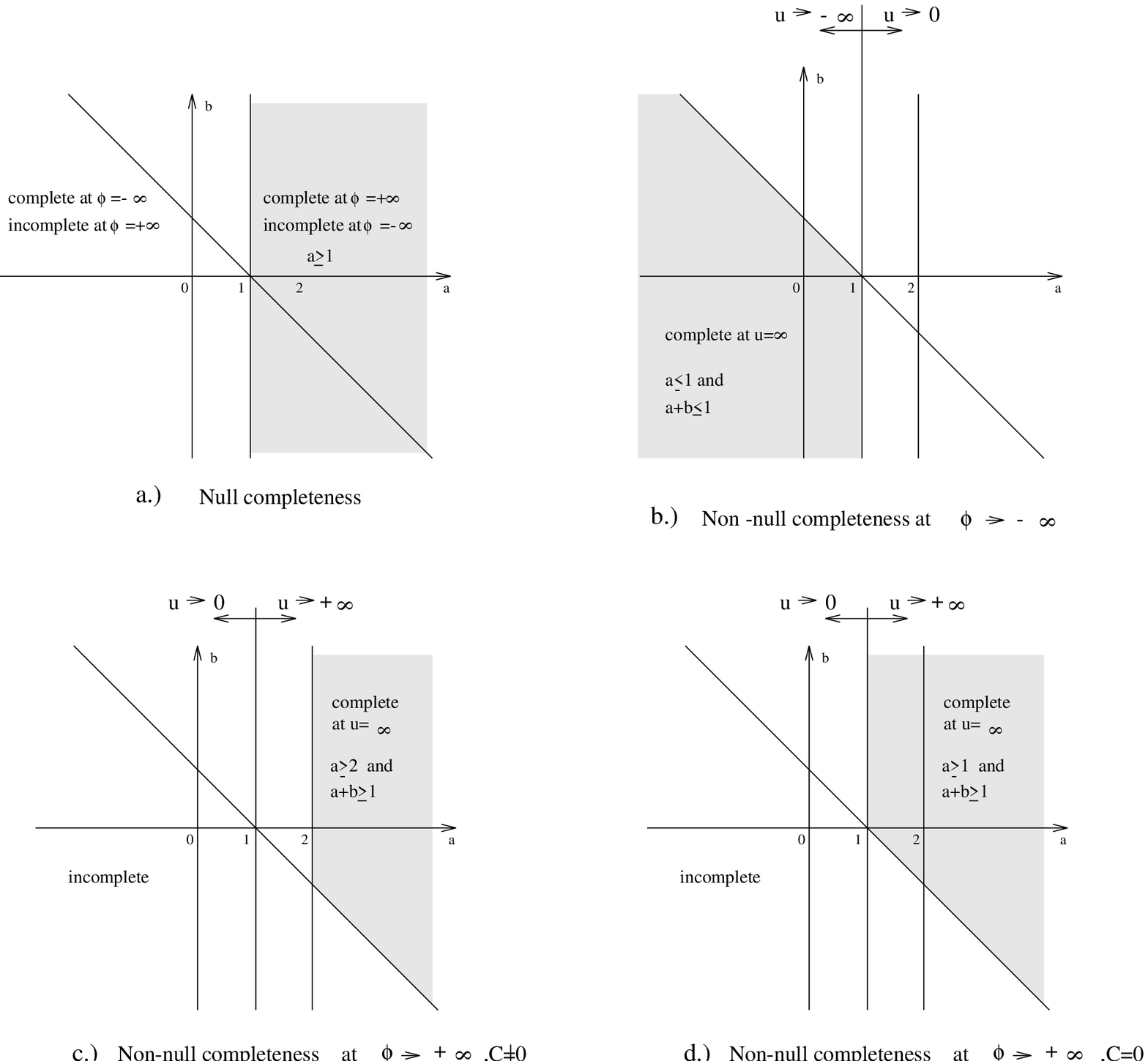}
\\
\centerline{Fig.\refstepcounter{figure}$\;$ \thefigure
\smallskip {}~~~Completeness of extremals}
\end{center}

\begin{center}
\leavevmode
\epsfxsize=12cm
\epsfbox{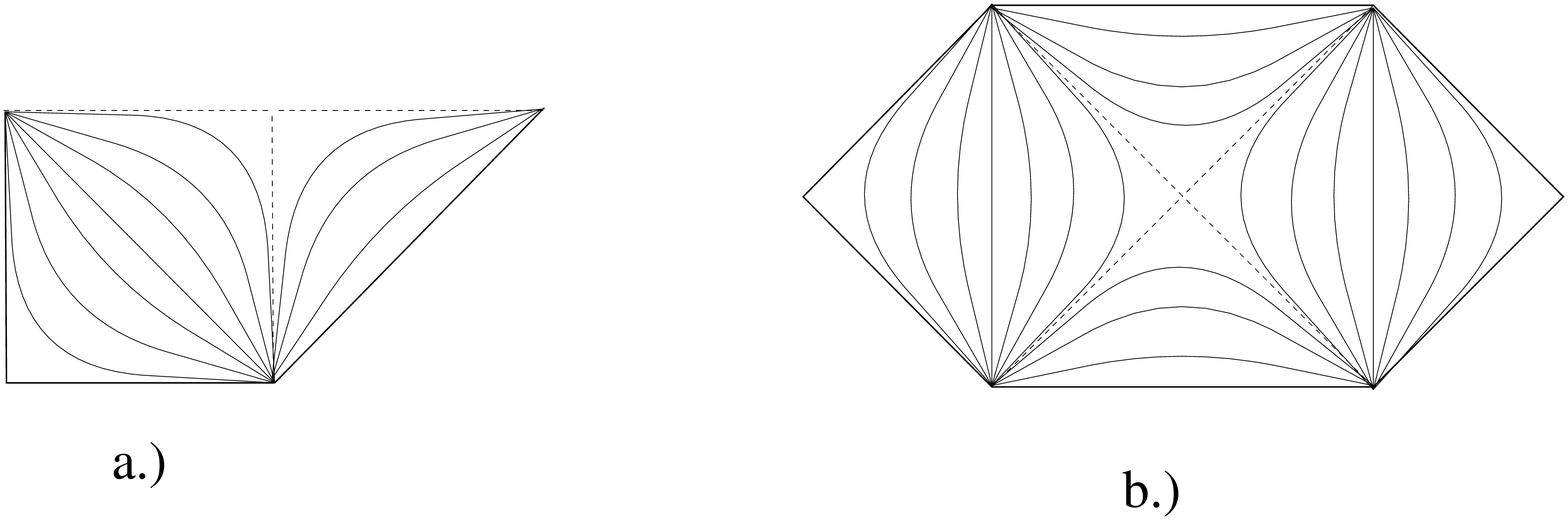}
\\
\centerline{Fig.\refstepcounter{figure}$\;$ \thefigure
\smallskip {}~~~a: Basic patch for a black hole; b: Generic black hole diagram}
\end{center}

\newpage

\begin{center}
\leavevmode
\epsfxsize=12cm
\epsfbox{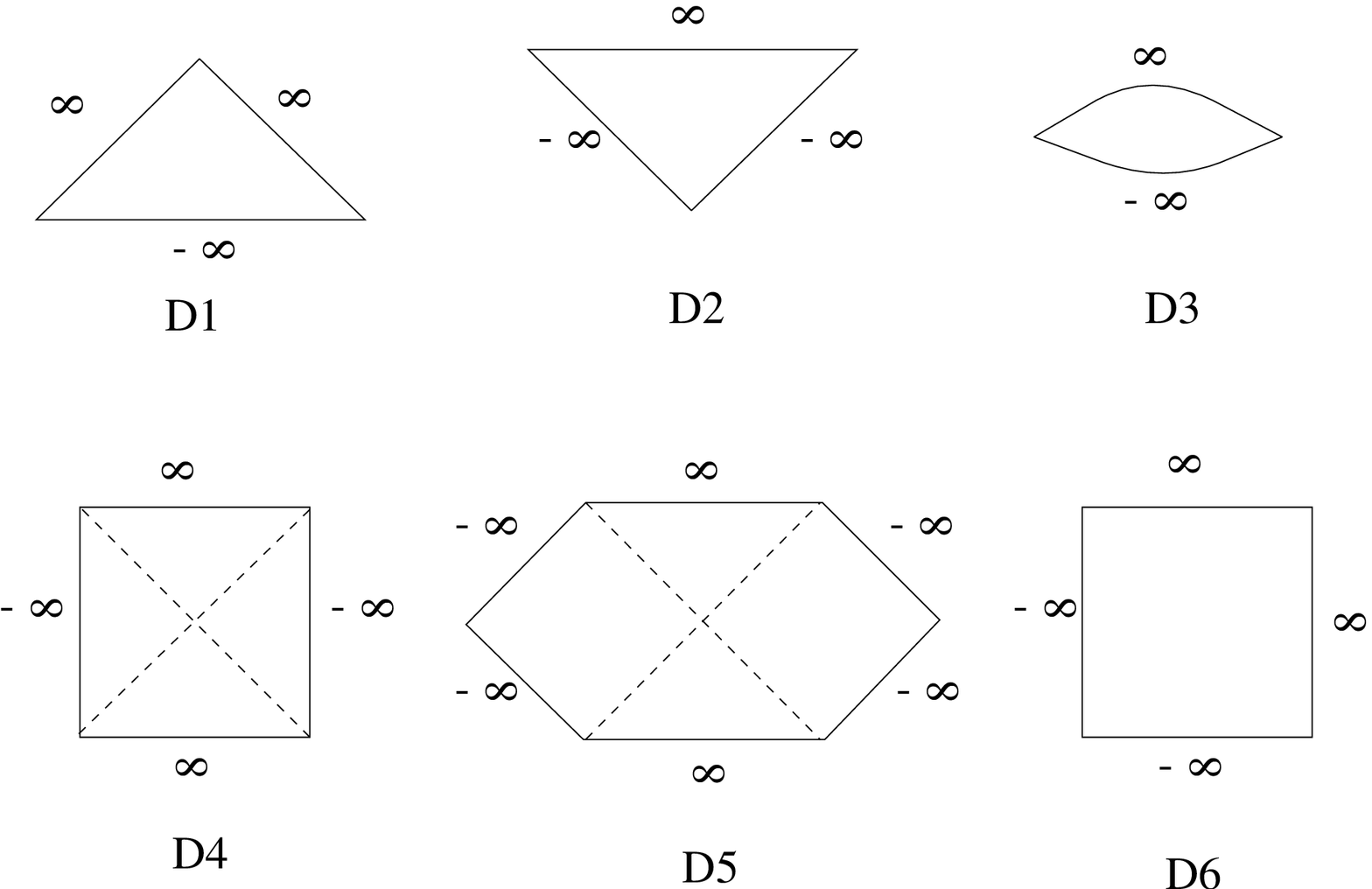}
\\
{Fig.\refstepcounter{figure}$\;$ \thefigure
\smallskip {}~~~Penrose diagrams for a generic black hole (\ref{ldil}). The 
boundaries correspond to infinite values of the dilaton field 
$\phi \to \pm \infty$ as indicated. Dashed lines inside a diagram denote 
horizons.} 
\end{center}

\begin{center}
\leavevmode
\epsfxsize=12cm
\epsfbox{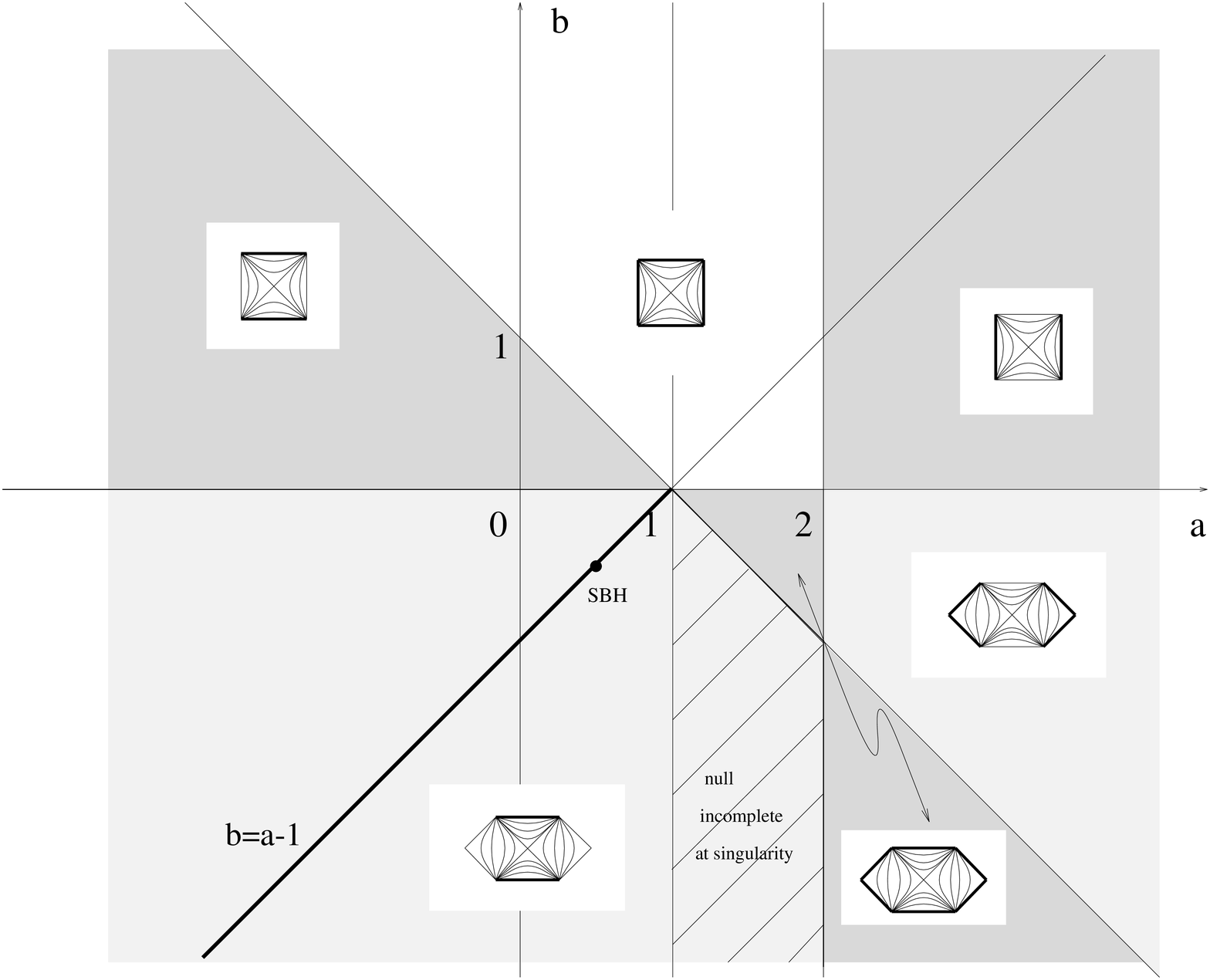}
\\
{Fig.\refstepcounter{figure}$\;$ \thefigure
\smallskip {}~~~Shape of the Penrose diagrams with horizon in dependence
of the values of the parameters $a$ and $b.$ The thick line $b=a-1$ indicates
the region of (\ref{future}).}
\end{center}


\begin{thebibliography}{99}
\bibitem[10]{very}A very comprehensive global analysis of generally 
		covariant 2d models, without concentrating especially on 
		dilaton theories, has only appeared recently \cite{alt}, whereas 
		models with torsion had been discussed in detail already 
		before \cite{kat93}.
\end{thebibliography}
\end{document}